\documentclass[12pt]{amsart}
\usepackage{amsmath}
\usepackage{amssymb}
\usepackage{mathrsfs}
\usepackage[a4paper]{geometry}
\theoremstyle{plain}
\newtheorem{teo}{Theorem}
\newtheorem{pro}{Proposition}
\newtheorem{lem}{Lemma}

\theoremstyle{definition}

\theoremstyle{remark}

\newtheorem*{nota}{Remark}
\newtheorem*{note}{Remarks}
\newcommand{\la}{\langle}
\newcommand{\ra}{\rangle}

\newcommand{\p}{\mathbb{P}}
\newcommand{\re}{\mathbb{R}}

\date{\today}
\title{Radial Dunkl Processes: Existence, uniqueness and hitting time} 
\begin{document}
\maketitle
\centerline{NIZAR DEMNI\footnote{SFB, Bielefeld university, e-mail: demni@math.uni-bielefeld.de }}  

\begin{abstract} 
We give shorter proofs of the following known results: the radial Dunkl process associated with a reduced system and a strictly positive multiplicity function is the unique strong solution for all times $t$ of a stochastic differential equation with a singular drift (see \cite{Scha} for the original proof and \cite{Chy} for a proof under an additional restriction),  the first hitting time of the Weyl chamber by a radial Dunkl process is finite almost surely for small values of the multiplicity function. Compared to the original proofs, ours give more information on the behaviour of the process. More precisely, the first proof allows to give a positive answer to a conjecture announced by Gallardo and Yor in \cite{Chy} while the second one shows that the process hits almost surely the wall corresponding to the simple root with a small multiplicity value.    
\end{abstract}

\section{Preliminaries}
To be self-contained, we begin by pointing out some facts on root systems and radial Dunkl processes. The reader is referred to \cite{Chy}, \cite{Hum}, \cite{Ros} for more details. 
Let $(V, \la,\ra)$ be a finite real Euclidean space of dimension $m$. A \emph{reduced} root system $R$ is a finite set of non zero vectors in $V$ such that : 
\begin{itemize}
\item[1] $R \cap \re \alpha = \{\alpha,-\alpha\}$ for all $\alpha \in R$, \\
\item[2] $\sigma_{\alpha}(R) = R$, 
\end{itemize} where $\sigma_{\alpha}$ is the reflection with respect to the hyperplane $H_{\alpha}$ orthogonal to $\alpha$: 
\begin{equation*}
\sigma_{\alpha} (x) =  x - 2\frac{\la\alpha, x \ra}{|\alpha|^2} \alpha,\, |\alpha|^2 := \la \alpha,\alpha\ra \quad x \in V.
 \end{equation*}
A simple system $S$ is  a basis of $\textrm{span}(R)$ which induces a total ordering in $R$. A  root $\alpha$ is positive if it is a positive linear combination of elements of $S$. The set of positive roots is called a positive system and is denoted by $R_+$. The (finite) reflection group $W$ is the group generated by all the reflections $\sigma_{\alpha}$ for $\alpha \in R$ and acts on $R$ via the relation (\cite{Hum})
\begin{equation*}
\sigma_{\alpha}\sigma_{\eta}\sigma_{\alpha} = \sigma_{\sigma_{\alpha}(\eta)}, \, \alpha,\eta \in R.
\end{equation*} 
Given a root system $R$  with positive and simple systems $R_+,S$,  define the {\it positive Weyl chamber} $C$ by: 
\begin{equation*}
C := \{x \in V, \, \langle \alpha , x \rangle > 0 \, \forall \, \alpha \in R_+\} = \{x \in V, \, \langle \alpha , x \rangle > 0 \, \forall \, \alpha \in S\} 
 \end{equation*} and $\overline{C},\partial C$ its closure and boundary respectively. 
The radial Dunkl process $X^W$ is defined as the $\overline{C}$-valued continuous paths Markov process whose generator is given by : 
\begin{equation*}
\mathscr{L}_k^Wu(x) = \frac{1}{2}\Delta u(x) + \sum_{\alpha \in R_+}k(\alpha)\frac{\langle \alpha,\nabla u(x)\rangle}{\langle\alpha, x\rangle} 
\end{equation*} 
where $u \in C^2(\overline{C})$ satisfies the boundary conditions $\langle\nabla u(x), \alpha\rangle = 0$ for all $x \in H_{\alpha},\, \alpha \in R_+$, and $k(\alpha) \geq  0 $ is a multiplicity function (a $W$-invariant function).  

\section{Motivation}
In order to motivate the reader and prepare for the first result, we exhibit some known examples. The first and easiest one corresponds to $V = \re, R = B_1 = \{\pm 1\}$. There is only one orbit so that $k(\alpha) := k \geq 0$ and $X^W$ is a \emph{Bessel} process (\cite{Rev}) of \emph{index} $\nu = k - 1/2$. When $k > 0$ and $X_0^W= x \geq 0$, it is the unique strong solution of the following stochastic differential equation with singular drift: 
\begin{equation*}
dX^W_t = dB_t + \frac{k}{X_t^W}dt, \quad t \geq 0,
\end{equation*}
where $B$ is a standard Brwonian motion. A multivariate well known example is given by the so-called $A_{m-1}$-type root system  defined by:
\begin{equation*}
R = A_{m-1} = \{\pm (e_i - e_j), \, 1 \leq i < j \leq m-1\}, 
\end{equation*} 
with positive and simple systems given by : 
\begin{equation*}
R_+ = \{e_i - e_j, \,1\leq i < j \leq m \}, \quad S = \{e_i - e_{i+1}, \,1\leq i  \leq m-1 \},
\end{equation*}
where $(e_i)_{1 \leq i \leq m}$ is the canonical basis of $\re^m$. In this case, $V = \re^m$, the span of $R$ is the hyperplane of $\re^m$ consisting of vectors whose coordinates sum to zero and $C = \{x \in \re^m, x_1 > \dots > x_m\}$. Besides, there is only one orbit so that $k(\alpha) := k \geq 0$ and $X^W = (X^{W,i})_{1 \leq i \leq m}$ satisfies :  
\begin{equation} 
\label{Khil}
dX_t^{W,i} = d\nu_t^i + k \sum_{j \neq i}\frac{dt}{X_t^{W,i} - X_t^{W,j}} \quad 1 \leq i \leq m , \quad t < \tau
\end{equation}
with $X_0^{W,1} > \dots > X_0^{W,m}$, where $(\nu^i)_i$ are independent BMs and $\tau$, \emph{the first collision time}, is defined by
\begin{equation*}
\tau := \inf\{t, \, X_t^{W,i} = X_t^{W,j},\, \textrm{for \, some} \, (i,j)\}.
\end{equation*}
This process was deeply studied in \cite{Cepa} (see also \cite{Cepa1}) and it was shown that (\ref{Khil}) has a unique strong solution for all $t \geq 0$ when $X_0^{W,1} \geq  \dots \geq X_0^{W,m}$ provided that $k > 0$. Then, Schapira proved in \cite{Scha} that the radial Dunkl process is the unique strong solution of 
\begin{equation*}
dX_t^W = dB_t  - \nabla \Phi(X_t^W) dt , \quad X_0 \in \overline{C} 
\end{equation*}
where $\Phi(x) = -\sum_{\alpha \in R_+}k(\alpha)\ln(\langle\alpha,x\rangle)$ subject to $k(\alpha) > 0$ for all $\alpha \in R$. Schapira's proof relies on a similar existence and uniqueness result for radial Heckman-Opdam processes and an auxiliary result showing that radial Dunkl processes are limiting (in law) processes of rescaled radial Heckman-Opdam processes. Meanwhile and independently, Chybiryakov (see \cite{Chy} p.170) provides another more restrictive proof based on martingale problems however assuming that the simple system is a basis of $V$ (which is no more valid for $R=A_{m-1}$). Since both proofs are long, we thought it is interesting to write a relatively short proof and this is achieved relying on an important result due C\'epa and L\'epingle on stochastic differential equations with singular drifts (\cite{Cepa}).  
Another result with a lengthy proof too states that the first hitting time $T_0$ of $\partial C$
\begin{equation*}
T_0 := \inf\{t > 0, X_t^W \in \partial C\}
\end{equation*} 
is finite almost surely when $0 \leq k(\alpha) < 1/2$ for at least one simple root $\alpha \in R_+$.  
The proof displayed in \cite{Chy} uses local martingales and is not precise in the sense that it does not indicate which wall does the process hit. That is why, we generalized the proof given in \cite{Cepa} for $A$-type root systems ($T_0 = \tau$ in this setting) to claim that, for such a simple root $\alpha$ in an arbitrary reduced root system, the one dimensional process $\la\alpha,X^W\ra$ hits zero almost surely, implying the finiteness of $T_0$. The paper contains also some consequences of the first result (see Theorem \ref{exi} below): we actually give a positive answer to a conjecture announced by Gallardo and Yor claiming that the size of the jumps performed by a Dunkl process $X$ up to a fixed time $t$ is almost surely finite for any starting point $X_0 = x \in V$. We also improve some known results on Wishart and Laguerre processes. This is done by relating the eigenvalues process of these matrix-valued processes to a radial Dunkl process associtaed with a $B$-type root system.      

\section{Radial Dunkl Process : Existence and Uniqueness of a strong solution} 
\begin{teo}
\label{exi}
Let $R$ be a reduced root system and recall that:
\begin{equation*}
\Phi(x) = -\sum_{\alpha \in R_+}k(\alpha)\ln(\langle \alpha,x\rangle) := \sum_{\alpha \in R_+}k(\alpha)\theta(\langle\alpha,x\rangle), \quad x \in C, 
\end{equation*}
where $k(\alpha) > 0$ for all $\alpha \in R_+$. Then $X^W$ is the unique strong solution of 
\begin{equation}\label{DE}
dY_t = dB_t - \nabla \Phi(Y_t) dt,  \quad Y_0 \in \overline{C}, \, t \geq 0, 
\end{equation}
where $B$ is a Brownian motion in $V$ and $Y$ is a continuous $\overline{C}$-valued process. 
\end{teo}
{\it Proof}: From Theorem 2. 2 in \cite{Cepa1}, the SDE: 
\begin{equation}\label{DE1}
dY_t = dB_t - \nabla \Phi(Y_t) dt + n(Y_t)dL_t, \quad Y_0 \in \overline{C} 
\end{equation} 
where $n(x)$ belongs to the set of  unitary inward normal vectors to $C$ at $x \in V$ defined by 
\begin{equation}
\langle x - a, n(x) \rangle \, \leq 0, \quad a \in \overline{C}, 
\label{Ines}
\end{equation}  
and $L$ is the boundary process satisfying: 
\begin{equation*} 
dL_t = {\bf 1}_{\{Y_t \in \partial C\}} dL_t, 
\end{equation*} 
has a unique strong solution for all $t \geq 0$. 
Moreover: 
\begin{eqnarray}\mathbb{E}\left[ \int_0^T {\bf 1}_{\{Y_t \in \partial C\}} dt \right] & = & 0 \label{Slim},\\ 
\mathbb{E}\left[\int_0^T |\nabla \Phi(Y_t)| dt\right] & < & \infty \label{lotfi}
\end{eqnarray}  
for all $T > 0$. All what we need is to prove that $(L_t)_{t \geq 0}$ vanishes. To proceed, we need two Lemmas. 

\begin{lem} 
\label{Hamdi} 
Set $dG_t : = n(Y_t)dL_t$. Then, for all $\alpha \in R_+$,  
\begin{equation*} 
{\bf 1}_{\{\langle Y_t,\alpha\rangle = 0\}} \langle dG_t, \alpha\rangle = 0. 
\end{equation*}
\end{lem}
{\it Proof}: The proof is roughly an extension to arbitrary root systems of the one given in \cite{Cepa} for $R= A_{m-1}$ . In order to convince the reader, we provide an outline. 
On the one hand, the occupation density formula yields: 
\begin{equation*}
\int_0^{\infty}L_t^a(\langle \alpha,Y\rangle)|\theta^{'}|(a)da = \langle \alpha,\alpha\rangle  \int_0^t |\theta^{'}|(\langle\alpha,Y_s\rangle) ds
\end{equation*}
where $L_t^a(\langle \alpha,Y \rangle)$ is the local time up to time $t$ at the level $a \geq 0$ of the real continuous semimartingale $\langle \alpha, Y \rangle \geq 0$ (\cite{Rev}). 
On the other hand, the following inequality holds (instead of (2.5) in \cite{Cepa1}) for all $a \in C$: 
\begin{align*}
\langle \nabla &\Phi(x),x-a \rangle = \sum_{\alpha \in R_+}k(\alpha) \theta^{'}(\langle \alpha,x\rangle)\langle \alpha,x-a\rangle \\& 
\overset{(\star)}{\geq} \sum_{\alpha \in R_+}k(\alpha)[b_{\alpha}|\theta^{'}|(\langle\alpha,x\rangle) - c_{\alpha}\langle\alpha,x-a\rangle - d_{\alpha}] \\&
\geq \min_{\alpha \in R_+}(b_{\alpha}k(\alpha))\sum_{\alpha \in R_+}|\theta^{'}|(\langle\alpha,x\rangle) - |x-a|\sum_{\alpha \in R_+}k(\alpha){c_{\alpha}}|\alpha| - 
\sum_{\alpha \in R_+}k(\alpha)d_{\alpha}
\\& := A \sum_{\alpha \in R_+}|\theta^{'}|(\langle\alpha,x\rangle) - B|x-a| - C
\end{align*}
by Cauchy-Schwarz inequality, where in $(\star)$, we used equation (2.1) in \cite{Cepa1}: let $g$ be a convex $C^1$-function on an open convex set $D \subset \re^m$, then $\forall a \in D$, there exist
$b,c,d > 0$ such that for all $x \in D$: 
\begin{equation*}
\langle \nabla g(x),x-a \rangle  \, \geq \, b|\nabla g(x)| - c |x-a| - d. 
\end{equation*}
 Note also that $A > 0$ since $b_{\alpha}k(\alpha) > 0$ for all $\alpha \in R_+$. Then, the continuity of $Y$, (\ref{lotfi}) and the above inequality yield: 
\begin{equation*}
\int_0^t |\theta^{'}(\langle \alpha,Y_s \rangle)|ds< \infty
\end{equation*} which implies that:
\begin{equation*}
\int_0^{\infty}L_t^a(\langle \alpha,Y \rangle)|\theta^{'}(a)|da < \infty 
\end{equation*}
Thus, $L_t^0(\langle \alpha,Y\rangle) = 0$ since the function $a \mapsto |\theta^{'}(a)|$ is not integrable at $0$. 
The next step consists in using Tanaka formula to compute 
\begin{align*}
dZ_t & := d[\langle\alpha,Y_t\rangle - (\langle\alpha,Y_t\rangle)^+]
\\& = {\bf 1}_{\{\langle\alpha,Y_t\rangle = 0\}}\langle\alpha,dB_t \rangle  -  {\bf 1}_{\{\langle \alpha,Y_t\rangle = 0\}} \langle\alpha, \nabla \Phi(Y_t) \rangle dt +   {\bf 1}_{\{\langle 
\alpha,Y_t \rangle = 0\}}\langle \alpha, dG_t \rangle
\end{align*}
for $\alpha \in S$. It is obvious that the second term vanishes. The first vanishes too since it is a continuous local martingale with null bracket (occupation density formula). 
As $Y_t \in \overline{C}$, then $dZ_t = 0$ a.s. which gives the result. $\hfill \blacksquare$

\begin{lem}
\label{Abass}
Let $x \in \partial C$. Then $\langle n(x), \alpha \rangle \neq 0$ for some $\alpha \in S$ such that $\langle x,\alpha \rangle = 0$.
\end{lem}
{\it Proof}: assume that $\langle n(x),\alpha\rangle = 0$ for all $\alpha \in S$ such that $\langle x,\alpha\rangle = 0$. The idea is to find a strictly positive constant $\epsilon$ such that $x - \epsilon n(x) \in \overline{C}$ then use the definition of inward normal vectors (see \eqref{Ines} above) to conclude that $n(x) = 0$. Our assumption implies that $\langle x,\alpha \rangle > 0$ for all $\alpha \in S$ such that $\langle n(x),\alpha \rangle \neq 0$. If such simple roots do not exist, that is, $\langle n(x),\alpha\rangle = 0$ for all $\alpha \in S$, then $x - \epsilon n(x) \in \overline C$ for all $\epsilon > 0$. Otherwise, if $\langle n(x),\alpha \rangle < 0$ for these simple roots, then $x- \epsilon n(x) \in \overline{C}$ for all $\epsilon > 0$. 
Finally, if none of these conditions is satisfied, choose 
\begin{equation*}
0 < \epsilon < \min_{\langle x,\alpha \rangle > 0, \langle n(x),\alpha \rangle  > 0}\frac{\langle x,\alpha \rangle}{\langle n(x),\alpha\rangle},  
\end{equation*}
to see that $x - \epsilon n(x) \in \overline{C}$. Substituting $a = x -\epsilon n(x)$ in \eqref{Ines} for the three alternatives, it then follows that $n(x)$ is the null vector, contradiction. 
$\hfill \blacksquare$\\ 
Now we proceed to end the proof of Theorem \ref{exi}.  Lemma \ref{Abass} asserts that 
\begin{equation*}
\{Y_t \in \partial C\} \subset \cup_{\alpha \in S}\{\langle Y_t,\alpha\rangle = 0, \langle n(Y_t),\alpha\rangle \neq 0\}
\end{equation*}
for all $t$. It follows that 
\begin{align*}
0 &\leq L_t \leq \sum_{\alpha \in S} \int_0^t {\bf 1}_{\{\langle Y_s,\alpha\rangle = 0, \langle n(Y_s),\alpha\rangle \neq 0\}}dL_s
\\& = \sum_{\alpha \in S} \int_0^t \frac{1}{\langle n(Y_s),\alpha\rangle}{\bf 1}_{\{\langle Y_s,\alpha\rangle = 0, \langle n(Y_s),\alpha\rangle \neq 0\}}\langle n(Y_s),\alpha\rangle dL_s = 0
\end{align*}
by Lemma \ref{Hamdi}. $\hfill \blacksquare$ 
\begin{note} 
1/When $m=1$, $(X_t)_{t \geq 0}$ is a Bessel process of dimension $\delta = 2k_0 + 1$ and  $k_0 > 0 \Leftrightarrow \delta > 1$. It is well known that the local time vanishes 
(see Ch. XI in \cite{Rev}).\\
2/Since the boundary $\partial C$ of $C$ is a cone, then for $x \in \partial C$, one has $cx \in \partial C$ for any $c \geq 0$. Letting $a = 0$ and $a = cx$ for $c > 1$, 
one gets from \eqref{Ines} that $\langle n(x),x\rangle = 0$ so that $n(x) \in x^{\bot}$ and $\la a,n(x) \ra \geq 0$. Moreover,  if $X_t \in H_{\alpha}$ for one and only one 
$\alpha  \in S$, then  $\la n(X_t),\alpha \ra \neq 0$ by Lemma \ref{Abass}  which together with Lemma \ref{Hamdi} yield 
\begin{equation*}   
{\bf 1}_{\{\langle Y_t,\alpha\rangle  = 0\}} \langle dG_t, \alpha\rangle = {\bf 1}_{\{\langle Y_t,\alpha\rangle = 0\}} \la n(X_t), \alpha \ra dL_t = 0. 
\end{equation*}
Hence, $(L_t)_{t \geq 0}$ vanishes. 
\end{note}

\section{Consequences}
\subsection{On a conjecture by Gallardo-Yor}
Recall that the Dunkl process $X$ is a $V$-valued Markov process with jumps whose projection on the orbits space, identified with $\overline{C}$, is $X^W$. Recall also that a jump at time $t$ can only occur in a direction of a root $\alpha$ provided that 
$X_t = \sigma_{\alpha}(X_{t-}) \neq X_{t-}$ so that the  jump's size is given by 
\begin{equation*}
\Delta X_t := X_t - X_{t-} = 2\frac{\la \alpha, X_{t-}\ra}{\la \alpha,\alpha \ra} \alpha.
\end{equation*}   
Then, it is conjectured that for a strictly positive multiplicity function (see \cite{Chy} p.127)
\begin{equation}\label{Dis1}
\sum_{s \leq t} |\Delta X_s|  = \sqrt{2}\sum_{s \leq t} \sum_{\alpha \in R_+} |\la \alpha, X_{s-}\ra| {\bf 1}_{\{X_s = \sigma_{\alpha} X_{s-} \neq X_{s-}\}}\, < \, \infty   
\end{equation}
almost surely for any $x \in V$, where we assumed without loss of generality that $|\alpha|^2 = 2$. The conjecture was shown in \cite{Chy} to be true for $x \in V \setminus \{0\}$ and uses  tedious computations together with the Markov property. Here, we use Theorem \ref{exi} to give a quick proof to the validity of the conjecture. The strategy consists in carrying the conjecture to the $W$-invariant setting. We start by recalling that after compensating the discontinuous function displayed in \eqref{Dis1} using the L\'evy kernel (\cite{Chy} p.123), it suffices to prove that  
\begin{align*}
\int_0^t ds \sum_{\alpha \in R_+}\frac{k(\alpha)}{|\la \alpha,X_{s} \ra|}
\end{align*}
has finite expectation for any starting point $x \in V$. To proceed, recall that the semi group density of $X$ is given for $(x,y) \in V^2$ by: 
\begin{equation*}
p_t^k(x,y) = \frac{1}{c_kt^{\gamma + n/2}} e^{-(|x|^2 + |y|^2)/(2t)} D_k\left(\frac{x}{\sqrt t}, \frac{y}{\sqrt t}\right) \omega_k(y)
\end{equation*} 
where 
\begin{equation*}
\gamma := \sum_{\alpha \in R_+} k(\alpha),\quad \omega_k(y) := \prod_{\alpha \in R_+}|\la \alpha,y\ra|^{2k(\alpha)}.
\end{equation*}
Recall also that any $y \in V$ is conjugated to one and only one element, say $y'$, belonging to $\overline{C}$. This gives the decomposition 
\begin{equation*}
V = \cup_{w \in W} w\overline{C}.
\end{equation*}
It follows that 
\begin{align*}
\mathbb{E}_x\left[ \sum_{\alpha \in R_+} \frac{k(\alpha)}{|\la \alpha,X_{s} \ra|}\right] &=  \frac{1}{2}\int_V p_s^k(x,y) \sum_{\alpha \in R} \frac{k(\alpha)}{|\la \alpha,y \ra|}dy 
\\& = \frac{1}{2}\sum_{w \in W}\int_{wC} p_s^k(x,y) \sum_{\alpha \in R} \frac{k(\alpha)}{|\la \alpha,y \ra|}dy 
\\& = \frac{1}{2}\int_C\sum_{w \in W} p_s^k(x,wy) \sum_{\alpha \in R} \frac{k(\alpha)}{|\la \alpha,wy \ra|}dy. 
\end{align*}
Now, from the very definition of root systems, one has $wR = R$ yielding 
\begin{equation*}
\sum_{\alpha \in  R}\frac{k(\alpha)}{|\la \alpha,wy \ra|} = \sum_{\alpha \in R}\frac{k(\alpha)}{|\la \alpha,y \ra|} = 2\sum_{\alpha \in R_+}\frac{k(\alpha)}{|\la \alpha,y \ra|} = 
2\sum_{\alpha \in R_+}\frac{k(\alpha)}{\la \alpha,y \ra}
\end{equation*}
for $y \in C$. Besides, if $ x = w_xx', \,x \in V, w_x \in W, x' \in \overline{C}$, then for $y \in C$ 
\begin{align*}
\sum_{w \in W} p_s^k(x,wy) &= \frac{1}{c_ks^{\gamma + n/2}} e^{-(|x|^2 + |y|^2)/(2s)} D_k^W\left(\frac{x}{\sqrt s}, \frac{y}{\sqrt s}\right) \prod_{\alpha \in R_+}|\la \alpha, wy\ra|^{2k(\alpha)}
\\& = \frac{1}{c_ks^{\gamma + n/2}} e^{-(|x|^2 + |y|^2)/(2s)} D_k^W\left(\frac{x'}{\sqrt s}, \frac{y}{\sqrt s}\right)\prod_{\alpha \in R}|\la \alpha, y\ra|^{k(\alpha)} 
\\& = \frac{1}{c_ks^{\gamma + n/2}} e^{-(|x|^2 + |y|^2)/(2s)} D_k^W\left(\frac{x'}{\sqrt s}, \frac{y}{\sqrt s}\right)\prod_{\alpha \in R_+}\la \alpha, y\ra^{2k(\alpha)} 
\end{align*}
where 
\begin{equation*}
D_k^W(x,y) := \sum_{w \in W}D_k(x,wy) = D_k^W(x',y)
\end{equation*}
is the generalized Bessel function (\cite{Chy}). Thus, we showed that

\begin{align*}
\int_0^t ds \mathbb{E}_x\left[ \sum_{\alpha \in R_+} \frac{k(\alpha)}{|\la \alpha,X_{s} \ra|}\right] &=  
\int_0^t ds\,\mathbb{E}_{x'}\left[\sum_{\alpha \in R_+} \frac{k(\alpha)}{\la \alpha,X_{s}^W \ra}\right]  
\\&=\int_0^tds \mathbb{E}_{x'}\left[\sum_{\alpha \in R_+}k(\alpha)|\theta'|\left(\la \alpha,X_s^W \ra\right)\right]
\end{align*} 
(recall that $\theta(u) := - \ln (u), u > 0$). A slight modification of the proof of Lemma \ref{Hamdi}  gives the inequality 
\begin{align*}
A\sum_{\alpha \in R_+} k(\alpha) |\theta'|(\la \alpha,y \ra) \leq \la \nabla \Phi (y), y - a \ra + B|y-a| + D 
\end{align*}
for all $y,a \in C$ for strictly positive constants $A,B,D > 0$. From the definition of $\Phi$, it is easy to see that 
\begin{equation*}
\la \nabla \Phi (y), y \ra = -\sum_{\alpha \in R_+} k(\alpha) = -\gamma. 
 \end{equation*}
Using the Cauchy-Schwartz inequality, it follows that 
\begin{equation*}
A\sum_{\alpha \in R_+} k(\alpha) \left|\theta'(\la \alpha,y \ra)\right| \leq  (B|a| + D +\gamma) + |a| |\nabla \Phi (y)| + B|y|.
\end{equation*}
Finally, if $k(\alpha) > 0$ for all $\alpha \in R_+$, then \eqref{Slim} holds and we already know that $|X^W|$ is a Bessel process of index $\gamma +n/2-1$  (\cite{Chy}) so that
\begin{equation*}
\int_0^tds \mathbb{E}_{x'}[|X_s^W|]  <\, \infty.
\end{equation*}
Therefore, the conjecture is proved. $\hfill \blacksquare$   

\begin{nota}
The proof of the above Corollary shows that  
\begin{equation*}
\mathbb{E}_x\left[\frac{1}{\la \alpha, X_1 \ra}\right] < \infty
\end{equation*} 
for all $\alpha \in R_+$. With regard to $p_1^k(x,y)$ and since the singularities of the function $y \mapsto 1/|\la \alpha,y \ra|$ lie only on the hyperplane orthogonal to $\alpha$, then one wants to claim that the Dunkl kernel $y \mapsto D_k(x,y)$ bahaves near to that hyperplane mostly as $1/|\la \alpha,y\ra|$ for fixed $x \in V$. However, we were not able to come rigorously to such quite important estimation near to hyperplanes.  
\end{nota}

\subsection{$\beta$-Laguerre processes and $B$-type root systems.} 
The $B_m$-type root system is defined by 
\begin{equation*}
R = \{\pm e_i,1 \leq i \leq m,\, \pm e_i \pm e_j,\, 1 \leq i < j \leq m\}.
\end{equation*}
Its positive and simple systems are given by 
\begin{equation*}
R_+ = \{ e_i,1 \leq i \leq m,\, e_i \pm e_j,\, 1 \leq i < j \leq m\}, \, S = \{ e_i - e_{i+1},\, 1 \leq i \leq m-1,\, e_m\}.
\end{equation*}
While we already pointed out that the radial Dunkl process of type $A$ fits the eigenvalues process of the symmetric and the Hermitian Brownian motion (\cite{Dyson}), the $B$-type root system turns out to be related to the eigenvalues process of Wishart and Laguerre processes (see \cite{Bru},\cite{Dem}) which is the unique strong solution of: 
\begin{equation}
\label{Lag}
d\lambda_i(t) = 2\sqrt{\lambda_i(t)}\, d\nu_i(t) + \beta\left[ \delta + \sum_{k \neq i}\frac{\lambda_i(t) + \lambda_k(t)}{\lambda_i(t) - \lambda_k(t)}\right]dt, \quad 1 \leq i \leq m 
\end{equation}
for $\beta = 1, 2$ and $\delta \geq m+1, m$ respectively, where $(\nu_i)_i$ are independent real Brownian motions and $\lambda_1(0) > \dots > \lambda_m(0) > 0$. The range of time is given by 
$t < \tau \wedge R_0$ where $\tau$ is the first collision time  and 
\begin{equation*}
R_0 := \inf\{t, \, \lambda_m (t) = 0\}
\end{equation*}
is the first hitting time of zero. It is known that $\tau = \infty$ almost surely and that $R_0 = \infty$ almost surely if $\delta \geq m+1,m$ respectively (\cite{Bru},\cite{Dem}). 

Define the $\beta$-Laguerre process as a solution, whenever it exists, of (\ref{Lag}) with arbitrary $\beta,\delta \geq 0$ up to $R_0 \wedge \tau$. 
Set $r_i := \sqrt{\lambda_i}$, then, for $t <  \tau \wedge R_0$: 
\begin{align*}
dr_i(t) &= d\nu_i(t) + \frac{1}{2r_i(t)}\left[\beta \delta -1 + \beta \sum_{j \neq i}\frac{r_i^2 + r_j^2}{r_i^2-r_j^2}\right] dt \\&
= d\nu_i(t) + \frac{\beta (\delta-m+1)-1}{2r_i(t)}dt + \frac{\beta}{2} \sum_{j \neq i}\left[\frac{1}{r_i(t) - r_j(t)} + \frac{1}{r_i(t) + r_j(t)}\right] dt \\&
= d\nu_i(t) + \frac{k_0}{r_i(t)}dt + k_1 \sum_{j \neq i}\left[\frac{1}{r_i(t) - r_j(t)} + \frac{1}{r_i(t) + r_j(t)}\right] dt
\end{align*}
with $2k_0 = \beta(\delta - m + 1) -1,\, 2k_1 = \beta$. Consequently, the process $r = (r_1,\dots,r_m)$ is a $B_m$-radial Dunkl process so that Theorem \ref{exi} asserts  that 
$(r(t))_{t \geq 0}$ is the unique strong solution for all $t \geq 0$ of (\ref{Lag}) provided that $k_0, \, k_1 > 0$, that is, $\beta > 0, \delta > m-1 + (1/\beta)$. 
This improves results from matrix theory (\cite{Bru}, \cite{Dem}): for Wishart processes, take $\beta =1$ so that the strong uniqueness for the eigenvalues process holds for all $\delta > m$. For Laguerre processes, take $\beta = 2$ so that strong uniqueness holds for $\delta > m - 1/2$. 

\section{Finiteness of the first hitting time of the Weyl chamber}
Let $T_0 := \inf\{t > 0, X_t^W \in \partial C\}$ be the first hitting time of the Weyl chamber. In \cite{Cepa}, where $R= A_{m-1}$ and $T_0 = \tau = \inf\{t > 0,\, X_t^{W,i} = X_t^{W,j}\, \textrm{for some}\, (i,j)\}$, authors showed that $T_0 < \infty$ a.s. if $0 \leq k_1 < 1/2$. More generally, it was shown (see \cite{Chy} p.169) that if $0 \leq k(\alpha) < 1/2$ for some $\alpha \in S$, then $T_0 < \infty$ almost surely. Here we prove a more precise statement: 
\begin{pro}
\label{Sacha}
Let $\alpha_0 \in S$ and $T_{\alpha_0} := \inf\{t > 0,\, \langle \alpha_0,X_t^W\rangle = 0\}$ such that $T_0 = \inf_{\alpha_0 \in S}T_{\alpha_0}$. If $0 \leq  k(\alpha_0) < 1/2$, then  
$(\langle \alpha_0,X_t^W\rangle)_{t \geq 0}$ hits a.s. $0$. Therefore $T_0 < T_{\alpha_0} < \infty$ a.s.
 \end{pro}
{\it Proof}: Assume $k(\alpha) > 0$ for all $\alpha \in R$ and let $\alpha_0 \in S$. Our scheme is a generalization of the one used in \cite{Cepa}, thus we shall show that the process 
$<\alpha_0,X>$ is a.s. less than or equal to a Bessel process of dimension $2k(\alpha_0) +1 : = 2k_0+1$. The result follows from the fact that $2k(\alpha_0) + 1 < 2$ when $k(\alpha) < 1/2$ so that the Bessel process hits zero a.s.. Using (\ref{DE}), one has for all $t \geq 0$
\begin{align*}
d\langle \alpha_0, X_t^W\rangle &= |\alpha_0| d\gamma_t + \sum_{\alpha \in R_+}k(\alpha)\frac{\langle \alpha,\alpha_0\rangle}{\langle\alpha,X_t^W \rangle} dt \\&
= |\alpha_0|d\gamma_t + k_0\frac{||\alpha_0||^2}{\langle\alpha_0,X_t^W \rangle}dt + \sum_{\alpha \in R_+\setminus \alpha_0}k(\alpha)\frac{\langle \alpha,\alpha_0\rangle}{\langle \alpha,X_t^W \rangle} dt.
\end{align*}
Set 
\begin{equation*}
R =  \cup_{j=1}^p R^j,  
\end{equation*} 
where $R^j,\, 1 \leq j \leq p$ denote the conjugacy classes of $R$ under the $W$-action, then 
\begin{equation*}
R_+ = \cup_{i=1}^p R_+^j \end{equation*}
so that: 
\begin{equation*}   
d\la\alpha_0, X_t^W\ra   = |\alpha_0|d\gamma_t + k_0\frac{|\alpha_0|^2}{\langle \alpha_0,X_t^W \rangle}dt + \sum_{j=0}^p k_j\sum_{\alpha \in R_+^j\setminus \alpha_0}
\frac{\langle\alpha,\alpha_0\rangle}{\langle \alpha,X_t^W\rangle}dt.
\end{equation*}
For a conjugacy class $R^j$ and $\alpha \in R^j$, if $\langle \alpha, \alpha_0\rangle := a(\alpha) > 0$ then, it is easy to check that $\langle \sigma_0(\alpha), \alpha_0\rangle = -a(\alpha)$ where $\sigma_0$ is the reflection with respect to the orthogonal hyperplane $H_{\alpha_0}$. Note that $\sigma_0(\alpha)$ belongs to the same conjugacy class of $\alpha$ and that 
$\sigma_0(\alpha) \in R_+ \setminus \alpha_0$ for $\alpha \in R_+ \setminus \alpha_0$ (see Proposition 1. 4 in \cite{Hum}). Hence, 
\begin{equation*}
d\langle\alpha_0, X_t^W\rangle   = |\alpha_0|d\gamma_t + k_0\frac{|\alpha_0|^2}{\langle\alpha_0,X_t^W\rangle} dt- \sum_{j=0}^p k_j\sum_{\substack{\alpha \in R_+^j\setminus \alpha_0 \\  a(\alpha) >0}}\frac{a(\alpha)\langle \alpha - \sigma_0(\alpha), X_t^W \rangle}{\langle\alpha,X_t^W \rangle\, \langle \sigma_0(\alpha),X_t^W\rangle}dt. 
\end{equation*}
Furthermore, 
\begin{equation*}
\alpha  - \sigma_0(\alpha) = 2\frac{\langle \alpha,\alpha_0\rangle}{\langle \alpha_0, \alpha_0\rangle}\alpha_0 \quad \textrm{so that} \quad \langle\alpha  - \sigma_0(\alpha) , X_t^W \rangle
= 2 a(\alpha)\frac{\langle \alpha_0, X_t^W\rangle}{|\alpha_0|^2}.   
\end{equation*} 
Consequently: 
\begin{equation*}
d \langle\alpha_0, X_t^W\rangle  = |\alpha_0|d\gamma_t + k_0\frac{|\alpha_0|^2}{\langle \alpha_0,^W_t \rangle}dt  + F_t \,dt 
\end{equation*} 
where $F_t < 0$ on $\{T_{\alpha_0} = \infty\}$. Using the comparison Theorem in \cite{Kar} (Proposition 2. 18. p. 293 and Exercice 2. 19. p. 294), one claims that 
$\langle \alpha_0,X_t^W \rangle  \, \leq \, Y_{|\alpha_0|^2t}^x$ for all $t \geq 0$ on $\{T_{\alpha_0} = \infty\}$, where $Y^x$ is a Bessel process defined on the same probability space with respect to the same Brownian motion, of dimension $2k_0+1$ and starting at $Y_0 = x \geq \, \langle\alpha_0,X_0^W\rangle > 0$. This is not possible since a Bessel process of dimension $< 2$ hits $0$ a.s.  (\cite{Rev}, Chap. XI).  $\hfill \blacksquare$
\begin{note}
1/When one allows the multiplicity function $k$ to take zero values at some orbits, the SDE \eqref{DE} holds up to $t < T_0$ (Corollary 6.7 p. 169 in\cite{Chy}). Thus our result remains valid under this assumption. 
\\
2/{\bf Open question}: Given two simple roots $\alpha_1,\alpha_2 \in S$ such that 
\begin{equation*}
0 \leq k(\alpha_1) \neq k(\alpha_2) < 1/2,
\end{equation*} 
that is belonging to different orbits, we already know that $T_{\alpha_i} < \infty, i = 1,2$. Is is possible to compare $T_{\alpha_1}$ and $T_{\alpha_2}$? one way to so that is to seek two processes $R_1,R_2$ such that 
\begin{equation*}
\p( \langle \alpha_1,X_t^W\rangle \, < R_1(t) <\, \langle \alpha_2,X_t^W \rangle \,< R_2(t), \, \textrm{for all } t \geq 0) = 1
 \end{equation*} 
and $R_1,R_2$ hits zero almost surely.
\end{note}

\end{document}